\documentclass[aps,pra,amsmath,amsfonts,amssymb,twocolumn,groupedaddress,showpacs]{revtex4}
\usepackage{graphicx}

\begin{document}

\providecommand{\openone}{\leavevmode\hbox{\small1\kern-3.8pt\normalsize1}}

\def\pmx{\begin{pmatrix}}
\def\emx{\end{pmatrix}}
\def\bsq{\begin{subequations}}
\def\esq{\end{subequations}}
\def\bsmx{\begin{smallmatrix}}
\def\esmx{\end{smallmatrix}}
\def\ovw{\overrightarrow}
\def\be{\begin{eqnarray}}
\def\ee{\end{eqnarray}}
\def\bee{\begin{eqnarray*}}
\def\eee{\end{eqnarray*}}
\newcommand{\proj}[1]{ | #1 \kb  #1|}
\def\ovb{\overline}

\newtheorem{thm}{Theorem}
\newtheorem{cor}[thm]{Corollary}
\newtheorem{lemma}[thm]{Lemma}
\newtheorem{conj}[thm]{Conjecture}
\newtheorem{prop}[thm]{Proposition}
\newtheorem{defn}[thm]{Definition}
        \def\pf{\medbreak\noindent{\bf Proof:}\enspace}
     \def\rmk{\medbreak\noindent{\bf Remark:}\enspace}
        \def\lanbox{{$\, \vrule height 0.25cm width 0.25cm depth 0.05mm \,$}}
        \def\qed{{\bf QED}}
        \def\Iff{\Longleftrightarrow}
      \def\Imp{\Longrightarrow}
        \def\iff{\Leftrightarrow}
      \def\imp{\Rightarrow}
             \def\half{{\textstyle \frac{1}{2}}}
               \def\frth{{\textstyle \frac{1}{2}}}
 \def\tr{{\rm Tr} \, }
 \def\trp{{\rm Tr} }

 \def\1rt2{ \tfrac{1}{\sqrt{2}}}
 \newcommand{\norm}[1]{ \| #1  \|}

        \def\fm{\lfloor m \rfloor}
      \def\Hil{{\mathcal H}}
\def\wh{\widehat}
\def\wt{\widetilde}
\def\ds{\displaystyle}
 \def\ts{\textstyle}
\def\bra{\langle}
\def\ket{\rangle}
\def\kb{ \ket \bra }
\def\dg{\dagger}
\def\ot{\otimes}
\def\lraw{\leftrightarrow}
\def\raw{\rightarrow}
\def\whm{\wh{\mu}}
\def\whn{\wh{\nu}}
\def\half{{\textstyle \frac{1}{2}}}
\def\frth{{\textstyle \frac{1}{4}}}
\def\PE{ {\rm ProbErr} }
\def\MPE{ {\rm MinProbErr} }
\def\dtsig{ \cdot \vec \sigma}
\def\bx{{\mathbf{\bold x}}}
\def\bv{{\mathbf{\bold v}}}
\def\bw{{\mathbf{\bold w}}}
\def\bz{{\mathbf{\bold z}}}
\def\bT{{\mathbf{\bold T}}}
\def\bs{{\mathbf{\bold s}}}
\def\bt{{\mathbf{\bold t}}}
\def\b0{{\mathbf{\bold 0}}}
\def\rmT{{\rm T}}
\def\mm{\! - \! }
\def\pp{\! + \! }
\def\e{\epsilon \,}
\def\vp{\varphi}
\def\trump{ \prec_T}
\def\nl{\newline}
\def\nn{\nonumber}

\title{Qubit channels with small correlations}

\author{Filippo Caruso$^{1}$, Vittorio Giovannetti$^{1}$, Chiara
Macchiavello$^{2}$, Mary Beth Ruskai$^{3}$}

\affiliation{$^{1}$ NEST CNR-INFM \& Scuola Normale Superiore,
    Piazza dei Cavalieri 7, I-56126 Pisa, Italy \\
$^{2}$ Dip. di Fisica and CNISM ``A. Volta'', Unit\`a di
Pavia, Via Bassi 6, I-27100 Pavia, Italy \\
$^{3}$ Department of Mathematics, Tufts University, Medford, MA
02155, USA}

\begin{abstract}
We introduce a class of quantum channels with correlations acting
on pairs of qubits, where the correlation takes the form of a
shift operator onto a maximally entangled state. We optimise the
output purity and show that below a certain threshold  the optimum
is achieved by partially entangled states whose degree of
entanglement increases monotonically with the correlation
parameter. Above this threshold, the optimum is achieved by the
maximally  entangled state characterizing the shift. Although, a
full analysis can only be done for the $2$-norm, both numerical
and heuristic arguments indicate that this behavior and the
optimal inputs are independent of $p > 1$ when the optimal output
purity is measured using the $p$-norm.
\end{abstract}


 \maketitle

\section{Introduction}

In usual memoryless channel model successive uses of the
communication line have the same noise~\cite{BS} and can be
described as a simple tensor product of channels. Recently, there
has been some interest in studying the behavior of these channels
with correlations \cite{MP,MPV} since such channels might be
regarded as a small first step in studying the much more complex
issue of channels with memory~\cite{BM,KW,VG,CCMR,RSGM,GM}. They
can also describe multiple access channels.   Thus, one is led to
consider a scenario in which with probability $1 - \mu$ qubits
encounter only uncorrelated tensor product noise, but with
probability $\mu$ they experience correlated noise. This situation
can be modelled by a channel of the form
 \be  \label{corrchan}
 \Phi = (1- \mu) \Psi  \ot \Psi  +  \mu \Gamma_{\rm corr},
\ee where all the correlations are in  the map $ \Gamma_{\rm
corr}$, which need not be a channel itself. Several papers
\cite{MP,MPV} have studied specific examples of this type for
which there is a critical value $\mu_c$ below which the optimal
input is a product state and above which the optimal input is
maximally entangled. Recently, Daems \cite{D} showed that this is
always the case for a class of channels called ``product Pauli''.
This is not, however, the most general channel of the form
\eqref{corrchan}, even for a pair of qubits.

In this note, we consider a class of channels with correlations
which exhibit quite different behavior. The map we use for $
\Gamma_{\rm corr}$ in \eqref{corrchan} is extremely simple; it is
simply the non-unital  channel which maps every input to a fixed
maximally entangled state. We study the optimal output purity,
rather than the channel capacity. For our channels, the
entanglement of the output input increases continuously with $\mu$
until it reaches a critical $\mu_c$, after which the optimal input
is always achieved with a maximally entangled state.

Although one is ultimately interested in the effect of
correlations on various types of channel capacity, we consider
here only the optimal output purity. For channels with some
covariance properties~\cite{HOLEVOcov}, one can make an explicit
connection between the classical capacity~\cite{HSW} and the
optimal output purity as measured by the minimal output von
Neumann entropy.   However, that need not hold in general and is
not true for the channels considered here.     Moreover, the
conjecture in  \cite{KDP} about capacity achieved with maximally
entangled states depends on the precise form of the channels and
is not relevant here.

Roughly speaking, one expects inputs whose outputs are close to
pure states to be the least corrupted; however, one can  have very
noisy channels which map all inputs to a region close to a fixed
pure state with little correlation with the input. Nevertheless,
the optimal output purity as measured by either the minimal output
entropy \cite{KR0,Shor1} or the maximal output $p$-norm \cite{AHW}
is of some interest and has been studied extensively. This is, in
part, due to the Shor equivalence~\cite{Shor6} established between
the conjectured additivity of minimal output entropy and other
long-standing additivity conjectures. The hope that the additivity
of minimal output entropy could be proved by showing that the
maximal output  $p$-norm is multiplicative, at least near  $p
\approx 1$, as  conjectured in \cite{AHW}, was recently shattered
by the discovery \cite{HW} of counterexamples to the latter for
all $p > 1$.    Nevertheless, the additivity conjectures remain
open and even the multiplicativity conjecture is known to hold in
the region $1 < p \leq 2$ for certain classes of channels
\cite{AF,F,KING1,KING11} and in other situations for $p = 2$
\cite{KR3,Mich}.  Thus, the optimal output  purity remains an
object of some interest.

For  $p>1$,  the $p$-norm of a state $\gamma$ is given by the
expression
\begin{eqnarray}
\norm{\gamma}_p
\equiv \left[ \mbox{Tr} (\gamma^p) \right]^{1/p}
\label{ppuri}\;.
\end{eqnarray}
One sometimes uses instead, the  R\'{e}nyi entropy~\cite{Renyi}
\be S_p(\gamma)\equiv  \tfrac{1}{1-p} \ln \tr (\gamma^p)  =
\tfrac{p}{1-p} \ln \norm{\gamma}_p \;, \ee which is known to
converge to the usual von Neumann entropy as $p \raw 1$. The
maximal output $p$-norm of a  CPT map $\Phi$ denoted
$\nu_p(\Phi)$, is the supremum of $\norm{\Phi(\gamma)}_p$
  over all input density  matrices $\gamma$, i.e.,
\begin{eqnarray}  \label{pmax}
 \nu_p(\Phi) \equiv \sup_{\gamma} \;  \norm{\Phi(\gamma)}_p  ~.
\label{mapnorm}
\end{eqnarray}
The minimal output entropy and Renyi entropy are similarly defined
as \be   \label{entmin}
     S_p(\Phi) \equiv  \inf_{\gamma} S_p[\Phi(\gamma)]\;,
\ee
and it is natural to refer to states which achieve the optimum in
\eqref{pmax} or \eqref{entmin} as optimal inputs.

This paper is organized as follows. In Section~\ref{oop} we
describe the class of channels we study and show how covariance
properties can be used to reformulate the optimization problem. In
Section~\ref{sect:opt} we use unitary transformations to simplify
the problem, solve it exactly when $p = 2$, reduce the general
case to the analysis of a single parameter, and report numerical
work which supports the conclusion that the optimal inputs are
independent of $p$. In Section~\ref{sect:heur}, which can be
skipped on first reading, we analyze the behavior of the output
eigenvalues under certain small perturbations. Although this
analysis does not yield a proof for $p \neq 2$, it does  support
our conjectures and yield a proof for $p = \infty$. In
Section~\ref{sect:conc} we summarize our conclusions in the form
of both conjectures and theorems, and summarize the evidence for
the former. These can be reformulated as statements about the
trumping relation \cite{AN1,AN2,Daf,DK,JP} which plays an
important role in entanglement catalysis.

In addition to the isomorphism ${\bf C}_4 \simeq {\bf C}_2 \ot
{\bf C}_2$, there is also an isomorphism between vectors in  ${\bf
C}_4$ and matrices in $M_2$, and a straightforward way to make
this correspondence using the Pauli matrices. We describe this in
Appendix~\ref{app:entang}. Although the results are
straightforward and/or well-known, they play an important rule and
it is useful to describe them in a fixed notation.

\section{Channel definition and properties}     \label{oop}

Let $\Psi_{\lambda}$ denote the qubit depolarizing
channel~\cite{KING11} \be \Psi_\lambda(\gamma) =  (1 - \lambda)
\tfrac{1}{d} I  (\tr \gamma) + \lambda \gamma \ee with  $\lambda
\in [ -\tfrac{1}{3},1]$. From this we generate a correlated
two-qubit  channel of the form~(\ref{corrchan}) whose action  is
  \be  \label{genchan}
      \Phi_{\beta,\mu,\lambda}(R)  =
(1 - \mu) \big( \Psi_{ \lambda} \ot \Psi_{\lambda} \big)(R)
          + \mu \; ( \tr R)\; \proj{\beta} \;,
  \ee
where $ 0 \leq \mu \leq 1$  and $|\beta \ket$ is  a fixed
maximally entangled state of the two qubits. Although
\eqref{genchan} is well-defined for any matrix  in $M_4$, we are
interested in the case of density matrices, for which $R > 0$ and
$\tr R = 1$. We will exploit the  covariance property of the
depolarizing channel, i.e. \be \Psi_{ \lambda} (U \rho U^\dag) = U
\Psi_{\lambda}(\rho) U^\dag \;, \ee which holds for any unitary
$U$ matrix in $M_2$. We find a relationship between the channel
obtained using the Bell state $| \beta_0 \ket \equiv
\tfrac{1}{\sqrt{2}}  \big( |00 \ket + |11 \ket \big)$ and any
other maximally entangled state by observing \be   \label{chancov}
  \lefteqn{  \Phi_{\beta_0,\mu,\lambda}\big[ (U \ot V)R (U^\dag \ot V^\dag) \big] } \qquad  \nn \\
      & =  &  (U \ot V) \big[ (1 \mm \mu) (\Psi_{ \lambda} \ot  \Psi_{ \lambda})(R) +
           \mu   \proj{\beta} \,  \big]   (U^\dag \ot V^\dag)   \nn  \\
        & =  &  (U \ot V)  \Phi_{\beta,\mu,\lambda}(R)  (U^\dag \ot V^\dag)
\ee
where
\be    \label{newbeta}
     |\beta \ket =  (U^\dag \ot V^\dag)  |\beta_0 \ket = (I \ot V^\dag \ovb{U} ) |\beta_0 \ket
\ee and the second equality used \eqref{trans}. Since unitary
transformations do not affect  eigenvalues, \be    \label{normeq}
  \norm {   \Phi_{\beta_0,\mu,\lambda}\big[ (U \ot V)R (U^\dag \ot V^\dag) \big] }_p
      =     \norm {   \Phi_{\beta,\mu,\lambda}(R) }_p.
\ee
when $|\beta\ket$ is given by \eqref{newbeta}.    Because the $p$-norm is
convex, it suffices to consider the optimization \eqref{pmax} for pure states
$R = \proj{\psi}$.

We now define an equivalence relation on pure states $| \psi \ket
\in {\bf C^4}$ by
 \be
\label{enteq}
      |\psi_1 \ket \cong   |\psi_2 \ket   \Leftrightarrow    \exists  ~ \hbox{unitary} ~ U, V  ~
        \hbox{:}  ~~ U \ot V  |\psi_1 \ket =   |\psi_2 \ket \;
        . \nonumber \\
\ee
 Since all members of a given equivalence class are related by
local unitaries, they have the same entanglement. In the matrix
picture described in Appendix~\ref{app:entang}, each equivalence
class is characterized by its singular values. Moreover, we can
characterize each equivalence class by its entanglement as
measured by the entropy of its reduced density matrix (or by
replacing $h$ in \eqref{hdef} by another  function strictly
monotone on $[0,1]$). In particular, we find it useful to use the
so-called ``linear entropy'' $E = 2(1- \tr \gamma^2)$ in numerical
work.

We now let $|\wh{\psi} \ket$ denote an equivalence class or, more
properly, a representative of the class with properties to be
specified in the next section. Then it follows from \eqref{normeq}
that \be  \label{key}
     &\sup_{\psi}&    \norm {   \Phi_{\beta_0,\mu,\lambda} ( \proj{\psi})
     }_p \; \; \; \; \; \; \nonumber \\
        & = &   \sup_{\wh{\psi}} \sup_{U,V}
        \norm {   \Phi_{\beta_0,\mu,\lambda}\big[ (U \ot V) \proj{\wh{\psi}} (U^\dag \ot V^\dag) \big] }_p  \nn \\
        & = &  \sup_{\wh{\psi}}  \sup_{\beta}  \norm {   \Phi_{\beta,\mu,\lambda}(\proj{\wh{\psi}}) }_p
\ee
with $|\beta\ket$ maximally entangled.

The observations above allow us to draw several conclusions
\begin{itemize}

\item[a)]  First,   \eqref{normeq}  implies that $|\psi \ket$ is  an optimal input for
$ \Phi_{\beta,\mu,\lambda}$ if and only if  $U \ot V |\psi \ket$ is an optimal input for
$ \Phi_{\beta_0,\mu,\lambda}$.  Therefore,
\be \label{unitaryeq}
\nu_p (\Phi_{\beta, \mu,\lambda}) =  \nu_p (\Phi_{\beta_0, \mu,\lambda}) \;,
\ee
The same conclusion can be reached by reversing the
 roles of $|\beta_0 \ket$ and $ |\beta \ket $ in \eqref{key}.
Thus,  it is sufficient to study  $\Phi_{\beta_0, \mu,\lambda}$.

\item[b)] If the optimal $|\wh{\psi} \ket $  in the reformulation
\eqref{key} is unique, then the set of optimal inputs for
$\Phi_{\beta_0, \mu,\lambda}$ is   a subset of $\{ |\psi \ket  =
(U^\dag \ot V^\dag)  |\wh{\psi} \ket  :  U, V ~ \hbox{unitary}
\}$. Thus, we expect that, excluding some trivial cases ($\lambda
= 0 $ or $\mu =  1$), all optimal inputs for a given channel have
the same entanglement.

\item[c)]  The singlet state  $|\beta_2 \ket = \sigma_2 |\beta_0
\ket$ satisfies the covariance condition $U \ot U |\beta_2 \ket =
|\beta_2 \ket$ for all unitary $U \in M_2$. Therefore, if we
modify the equivalence relation \eqref{enteq} by restricting to $U
= V$,
 \be
  \nu_p (\Phi_{\beta_2, \mu,\lambda}) =
     \sup_{\widetilde{\psi} }    \norm {   \Phi_{\beta_2,\mu,\lambda} ( \proj{\widetilde{\psi}}) }_p
 \ee
where $|\widetilde{\psi} \ket$ denotes a representative of the
modified equivalence class.
 \end{itemize}

Before performing the optimization, we mention some trivial cases,
always using the channel with $|\beta_0 \ket$ for the shift. When
$\lambda = 0$ or $\mu = 1$, all inputs have the same output,
namely $\tfrac{1}{4} (1 - \mu) I + \mu \proj{\beta_0}$. We will
see that these are the only cases in which inputs with different
entanglement are optimal. When $\mu = 1$, the optimal  input
$|\beta_0 \ket$ is unique and the optimal output is also $|\beta_0
\ket$. Henceforth, we will assume that  $\lambda \in (0,1)$ and
$\mu \in [0,1)$. When $\mu = 0 $ every product state is an optimal
input; at the end of Section~\ref{sect:a1opt} we prove that, as
expected, all other states are non-optimal.

\section{Optimization}   \label{sect:opt}

\subsection{Simplifying the input}     \label{sect:newform}

After taking into account the normalization condition and
irrelevance of an overall phase factor, the optimization problem
for $\nu_p (\Phi_{\beta_0, \mu,\lambda})$ involves three complex,
or six real, variables. The covariance used to obtain \eqref{key}
reduces this to four real variables, one for $\wh{\psi}$ and three
for $|\beta \ket$. Moreover, the additional symmetry noted in (c)
above allows an immediate reduction to three real variables.  We
use a different approach; the reduction from four to three real
variables is obtained following \eqref{more}. In
Section~\ref{sect:a1opt}, we find that it would suffice to analyze
the dependence on a single variable with the others fixed.

Using the notation of Appendix~\ref{app:entang} for the maximally
entangled Bell states, we can write an  arbitrary state $\psi \in
{\bf C}_4$ as
  \be   \label{input}
       |\psi \ket = \sum_k a_k | \beta_k  \ket  = \sum_k a_k (I \ot \sigma_k)  | \beta_0  \ket =
         (I \ot  A )  | \beta_0  \ket \;, \; \; \; 
  \ee
where  $A =    \sum_k a_k  \sigma_k   \in M_2$.  Moreover, we can
use the SVD to choose unitary matrices $U ,V $ so that $ (U \ot V
) |\psi \ket =  (I \ot  V  A U^T)   | \beta_0  \ket $ with $V  A
U^T$ diagonal and positive, as discussed in
Appendix~\ref{app:entang}. It will be convenient to write the
corresponding state \eqref{psiUV} using an angular variable
$\theta \in [0, \tfrac{\pi}{2} ]$ so that \be   \label{theta}
 (U  \ot V ) |\psi \ket =
      | \psi_{\theta} \ket   &  \equiv  &  \1rt2( \cos \tfrac{\theta}{2} +   \sin \tfrac{\theta}{2}) |\beta_0 \ket \nonumber
      \\
      &+&  \1rt2( \cos \tfrac{\theta}{2} -  \sin \tfrac{\theta}{2}) |\beta_3 \ket \;,
\ee
which implies
 \be  \label{input-theta}
   | \psi_{\theta} \kb \psi_{\theta} | &=& \half \Big[ I \ot I + \sigma_z \ot \sigma_z + \cos \theta
       \big(\sigma_z \ot  I +  I \ot \sigma_z \big) \nonumber \\
       &+&
       \sin \theta    \big(\sigma_x \ot \sigma_x - \sigma_y \ot \sigma_y \big)\Big] \;.
 \ee

\subsection{Computing $\norm {   \Phi_{\beta,\mu,\lambda}( \proj{\psi_{\theta}}) }_p$}

Since $\Psi_{ \lambda}(\sigma_k)  = \lambda \, \sigma_k $  it is
straightforward to see that \eqref{input-theta} implies
 \be
\lefteqn{  \big( \Psi_{ \lambda} \ot \Psi_{\lambda} \big)
\big( | \psi_{\theta} \kb \psi_{\theta} | \big) }
  \\ & = &
    \tfrac{1}{4}\Big[ I \ot I +  \lambda^2 \sigma_z \ot \sigma_z + \cos \theta
     \lambda \big(\sigma_z \ot  I +  I \ot \sigma_z \big) \nonumber \\
     &+&
       \sin \theta   \lambda^2  \big(\sigma_x \ot \sigma_x - \sigma_y \ot \sigma_y \big) \Big] \nn \;,
 \ee
which can be written as
 \begin{widetext}
 \be    \label{bb}   \big( \Psi_{ \lambda} \ot \Psi_{\lambda} \big)
\big( | \psi_{\theta} \kb \psi_{\theta} | \big) = \frac{1}{4} \pmx
  1 + \lambda^2 \pp 2 \lambda \cos \theta  & 0 &  0 &
       2    \lambda^2  \sin \theta \\
       0 &   (1 \mm \lambda^2) & 0 & 0 \\
       0 & 0 &   (1 \mm \lambda^2) &  0 \\
 2    \lambda^2  \sin \theta  &0 & 0 &
 1+ \lambda^2 \mm 2 \lambda \cos \theta
      \emx.
\ee
Now writing  $V^\dag \ovb{U} = \pmx a & \ovb{b} \\ - b &
\ovb{a} \emx $ with $|a|^2 + |b|^2 = 1$, we find
 \be
\label{betaU}
     \bra \beta | = \bra \beta_0| (V^\dag \ovb{U})^\dag  =
        \bra \beta_0|  \pmx  \ovb{a} & -\ovb{b} \\ b & a \emx =
         \pmx \ovb{a} & - \ovb{b} & b & a \emx \;.
\ee
Using \eqref{bb} and \eqref{betaU} we find that
$ \Phi_{\beta,\mu,\lambda}( \proj{\psi_{\theta}})$  is given by
\bee
   \frac{1 \mm  \mu}{4} \pmx
  1 \pp \lambda^2 \pp 2 \lambda  \cos \theta  & 0 &  0 &
       2    \lambda^2  \sin \theta \\
       0 &    1 \mm \lambda^2  & 0 & 0 \\
       0 & 0 &    1 \mm \lambda^2  &  0 \\
 2    \lambda^2  \sin \theta  &0 & 0 &
 1\pp \lambda^2 \mm 2 \lambda  \cos \theta  \emx   + \frac{\mu}{2}
 \pmx    |a|^2 & - a \ovb{b} & ab & a^2 \\  -\ovb{a} b & |b|^2 & - b^2 & -ab \\
         \ovb{ab} & - \ovb{b}^2 & |b|^2 & a \ovb{b} \\  \ovb{a}^2 & - \ovb{ab} & \ovb{a} b & |a|^2
          \emx.
\eee
 It is evident that  $\pmx 0 & b & \ovb{b} & 0 \emx^T$ is an
eigenvector of both matrices above. This suggests that we make a
simplification exploiting  the block structure in the left matrix
using a basis which includes the known eigenvector. We act first
on both matrices with the  permutation matrix $P$ which exchanges
the 2nd and 4th rows and columns, and then make a unitary
transformation which preserves the block structure and achieves a
partial diagonalization.   Thus, we replace each matrix above by $
W^\dag P (~~) P W$  where \be    \label{more} P = \pmx 1 & 0 & 0 &
0 \\ 0 & 0 & 0 & 1 \\ 0 & 0 & 1 & 0 \\ 0 & 1 & 0 & 0 \emx  \qquad
     W = \1rt2 \pmx \frac{1}{|a|} \pmx  a & ~a \\  \ovb{a} & - \ovb{a}  \emx  & \hbox{{\Large 0 }} \\
                  \hbox{{\Large 0 }}   &  \frac{1}{|b|} \pmx   \ovb{b} &  \ovb{b}   \\  b & -b     \emx   \emx \;.
\ee
 Although these unitary transformations do {\em not} preserve
entanglement, they do not affect the eigenvalues of the output.
After introducing the shorthand $c_\vp \equiv \cos \vp , ~ s_\vp
\equiv  \sin \vp , S = 2 \lambda \sin \theta, C  = 2 \lambda \cos
\theta$, and $M_ {\mu}=  {4 \mu}/({1 - \mu})$,  we find that the
transformed output density matrix is
 \be \label{output}
   \frac{1 \mm \mu}{4}  \! \pmx  1 \pp  \lambda^2 +c_\vp   \,\lambda S +
    M_ {\mu}|a|^2 &
     C+ i  s_\vp    \,   \lambda S& 0 &  M_ {\mu}\sqrt{ |a|^2(1 - |a|^2) } \\
       C- i  s_\vp    \,   \lambda S&
        1 + \lambda^2 -c_\vp   \,\lambda S  & 0 & 0 \\
       0 &    0 &    1 \mm \lambda^2  & 0   \\
 M_{\mu}  \sqrt{ |a|^2(1 - |a|^2) } &    0 & 0 &    1 \mm \lambda^2 \pp  M_\mu (1 - |a|^2)  \emx    .
\ee
 \end{widetext}
We have now reduced the optimization to a
problem in three real variables, $\sin \theta,  \vp, |a|$.

\subsection{Optimization for $p = 2$ }  \label{sect:p2}

\begin{figure}
\begin{center}
 \includegraphics[width=8.5cm]{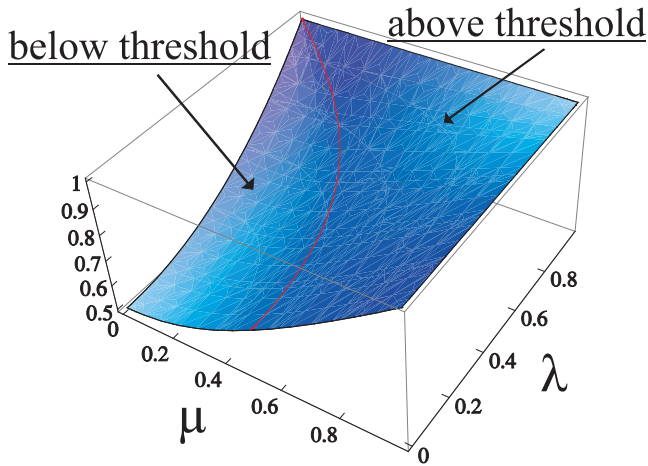}
\caption{Plot of the $2$-norm for the channel
$\Phi_{\beta_0,\mu,\lambda}$. The continuous curve represents the
threshold boundary. For $\mu=1$ or $\lambda=1$ the norm is $1$: in
the former case the channel sends every input state into
$\proj{\beta_0}$; in the latter, it transforms the input
$\proj{\psi}$ into a mixture of $\proj{\psi}$ and $\proj{\beta_0}$
so that choosing $|\psi\rangle = |\beta_0\rangle$ gives  a pure
output.}\label{fig1}
\end{center}
\end{figure}

It follows from  \eqref{output}  and some elementary algebra that
\be
\lefteqn{\norm {   \Phi_{\beta,\mu,\lambda}( \proj{\psi_{\theta}}) }_2^2} \label{pdue} \\
&=&  (\tfrac{1-\mu}{4})^2  \Big\{
 M_{\mu}^2 + 2 M_{\mu} \big[ 2 \lambda^2 |a|^2 + (1 \mm \lambda^2) + c_\vp \lambda S \big]
 \nonumber \\
 &+& 4(1 + \lambda^2)^2  - 2S^2( 1 - \lambda^2 ) \Big\}\nn
\ee
 Since only the coefficient of $M_ {\mu}$ includes any
dependence on $a$ and $M_ {\mu} > 0$,  it follows that the 2-norm
is largest when this coefficient is largest (see Fig. \ref{fig1}).
This occurs when $|a| = 1$, and $c_\vp  = 1$. After making these
choices  we find \be   \label{max2p} \norm{  ~~ }_2^2 & = &
 (\tfrac{1-\mu}{4})^2 \big[M_{\mu}^2 + 2M_{\mu} \big(1 + \lambda^2 +   \lambda S \big)
  \nn \\    &+& 4(1 + \lambda^4) + 2 (C^2 + \lambda^2 S^2)  \big] \nn \\
        & = &   (\tfrac{1-\mu}{4})^2 \big[M_{\mu}^2 + 2M_{\mu}
        \big(1 + \lambda^2 + 2 \lambda^2 \sin \theta \big)
      \nn \\ &+& 4(1 + \lambda^2)^2  - 4 \lambda^2( 1 - \lambda^2 ) \sin^2 \theta \big]
\ee
 which can be regarded as a quadratic function of $\sin \theta$
whose optimization is straightforward. Since we obtain an
equivalent problem whenever the optimum is achieved with $|a| =
1$, we give the details in the next section.

\subsection{Consequences of $|a| = 1$  optimal}
 \label{sect:a1opt}

We now describe the conclusions one can reach if the optimal
output $p$-norm is achieved when $|a| = 1$. We will show that when
this is the case, the optimal inputs are the same as for $p = 2$.

When $|a| = 1$, the output density matrix \eqref{output} is block
diagonal. Two of its eigenvalues  are $\frth (1-\mu)(1-
\lambda^2)$ and the remaining two are
 \be \label{evala1}
  && \frth (1-\mu)(1- \lambda^2) + \half \mu\\
   &\pm&   \frth (1-\mu)
       \sqrt{ \frth M_{\mu}^2 + 4 \lambda^2 - (1 - \lambda^2) S^2 + M_ {\mu}c_\vp \lambda S
       } \; . \nn
\ee The output $p$-norm will be largest, and output entropy
smallest, when the quantity under the square root is largest.
Therefore, one should choose $c_\vp = 1$. This implies that
$V^\dag \ovb{U} = I$ or, equivalently, that $V = \ovb{U} $.

\begin{figure}
\begin{center}
  \includegraphics[width=9cm]{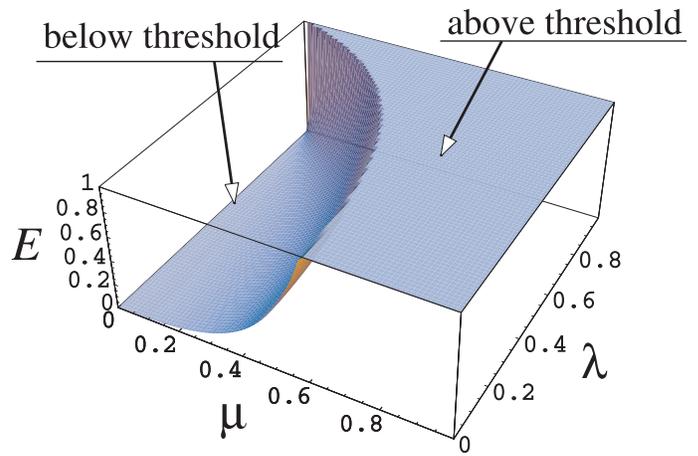}
\caption{Entanglement
  of the optimal input states which maximize the
$2$-norm for the map $\Phi_{\beta_0,\mu,\lambda}$. For $\mu \in
[0,\mu_c)$ the entanglement of  the optimal inputs is a strictly
monotone function of $\mu$; for $\mu \geq \mu_c$ the optimal input
is the maximally entangled state $|\beta_0 \ket$.} \label{fig2}
\end{center}
\end{figure}

The optimization problems for both  \eqref{max2p} and
\eqref{evala1} then reduce to maximizing a function of the form
 \be   \label{max11}
   -  4(1 - \lambda^2) \lambda^2 \sin^2 \theta + 2 \lambda^2 M_ {\mu  } \sin \theta + {\rm constant}
 \ee
which is quadratic in $ \sin \theta > 0$ when $\lambda \in (0,1)$.
This is largest when $ \sin \theta  = \min \{ 1,
\tfrac{\mu}{1-\mu} \tfrac{1}{1-\lambda^2}\}$. We distinguish two
situations characterized by the threshold value  $\mu_c \equiv
\frac{1 - \lambda^2}{2 - \lambda^2} $ corresponding to the
boundary  $\sin \theta =1$.
  \begin{itemize}
\item{\bf Below Threshold}:  If     $\mu < \mu_c$ then
  the maximum value of (\ref{max11}) is achieved for
  $4  \sin \theta (1 - \lambda^2) = M_\mu =  4 \tfrac{\mu}{1-\mu}$, or,
equivalently
\be   \label{thetaopt}
     \theta =   \sin^{-1} \left[ \frac{\mu}{1-\mu} \,  \frac{1}{1 - \lambda^2} \right] =
      \sin^{-1} \left[\frac{ \mu (1 - \mu_c)}{\mu_c ( 1 -\mu)}\right]  \;.
\ee
 Moreover,  since  $V = \ovb{U} $, the optimum is achieved with
a family of input states of the form
 \be
|\psi_{\rm opt}\ket   = (V^T \otimes V^\dag)  | \psi_\theta\ket \;,
\label{opt2}
 \ee
with $\theta$ as in \eqref{thetaopt}, $|\psi_\theta\ket$ as in
\eqref{theta}, and $V$ an arbitrary unitary.    Using
\eqref{reddm}, one finds that the reduced density matrix  of
$|\psi_\theta\ket$ is \be   \label{rdm.theta}
     \gamma_{\theta} = \half[ I +  \cos \theta  \, \sigma_ 3]
\ee
 and the entanglement of any optimal input is simply $h(\cos
\theta)$ where $h$ is the binary entropy \eqref{hdef}, or $E =
\sin^2 \theta$ when the linear entropy is used (see Fig.
\ref{fig2}).

\item{\bf At or above Threshold}:  For   $\mu \geq \mu_c$ the
maximum value of \eqref{max11} is achieved for $\sin \theta = 1$.
Then  \eqref{theta} gives $| \psi_\theta\ket =  |\beta_0 \ket $
and the diagonal matrix $D_\theta$   is simply the identity so
that $V^\dag D_{ \theta }V = V^\dag V = I$.  Thus, there is a
single optimal input, the maximally entangled state $|\beta_0
\ket$.
    \end{itemize}

Note that for $\mu = 0$ the value of $|a|$ is irrelevant and the
optimization gives a special case of \eqref{max11} with the
optimum achieved for $\theta = 0$, consistent with
\eqref{thetaopt}. Because there is no dependence on $a$  in this
case, this result holds for all $p$ and for all $|\beta \ket$ in
\eqref{key}. Hence,  $V^\dag \ovb{U}$ is arbitrary and we recover
the expected result that all product inputs are optimal. Because
$\theta \neq 0$ can never yield a state in the equivalence class
with no entanglement, we also find that a  state which is not a
product can not be optimal.

\subsection{Numerical results}   \label{sect:num}

\begin{figure}
\begin{center}
\includegraphics[width=7.5cm]{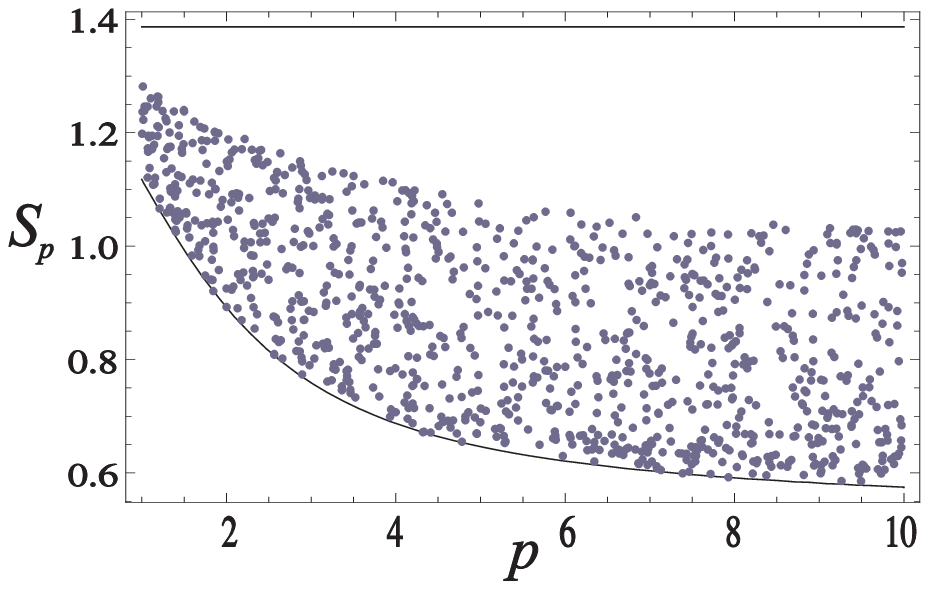}
\includegraphics[width=7.5cm]{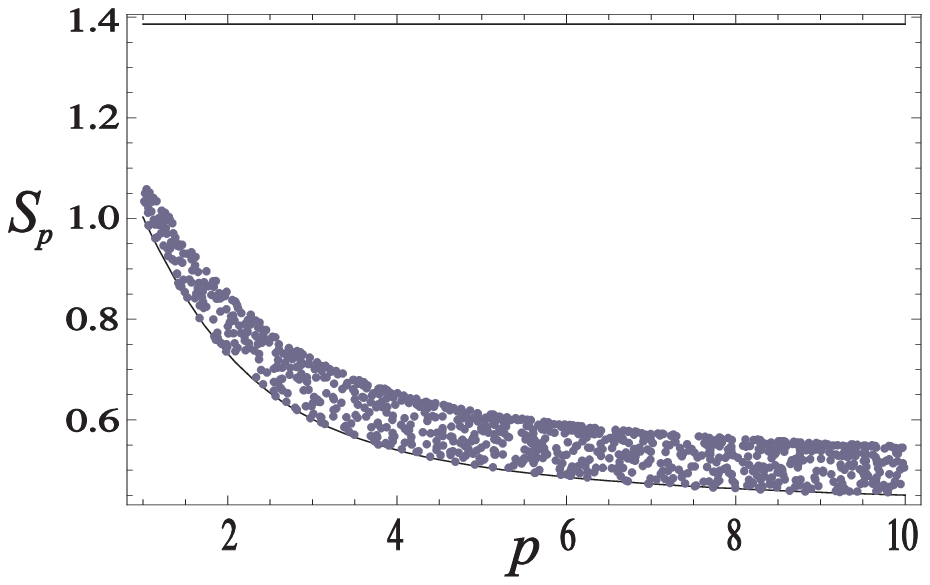}
   \end{center}
\caption{Scatter plots of the output Renyi entropy
$S_p[\Phi_{\beta_0,\mu, \lambda}(\proj{\psi})]$ as a  function of
$p$ for randomly chosen inputs compared to that for the
conjectured optimal input. The horizontal line corresponds to the
maximum possible Renyi entropy of $\ln 4$. Top: $ \lambda =
\tfrac{1}{2},  \mu = \tfrac{1}{4} <  \mu_c = \tfrac{3}{7}$ below
threshold. Bottom: $\lambda = \tfrac{1}{3}, \mu = \tfrac{1}{2}
> \mu_c = \tfrac{8}{17}$ above threshold.} \label{fig3}
\end{figure}

It is not easy to perform an exact analytical analysis of the
output $p$-norms for $p\neq 2$. However extensive numerical
studies of optimization were carried out using the equivalent
Renyi entropy \cite{Renyi}, for over 2000 pairs of randomly chosen
value of $\mu$ and $\lambda$. In all cases, we found that the
input states which are optimal for $p=2$ are also optimal for
$p>1$. In Fig.~\ref{fig3}, we show typical numerical results of
our findings by comparing the minimal R\'{e}nyi output
entropies~(\ref{entmin}) for randomly chosen inputs with that of
optimal inputs for $p = 2$, which lie on the bottom curve. In all
cases the output Renyi entropy was larger than the expected
minimum.

{\section{Partial analysis with heuristics}   \label{sect:heur}

\subsection{Eigenvalue behavior from characteristic polynomial}

In this section, we attempt to show that the optimal $p$-norm is
attained when $|a| = 1$ be showing that it increases monotonically
with $|a|$.    Although our argument is incomplete and can not
exclude fluctuations under some conditions, it does provide
additional support for this conjecture. To analyze the behavior of
the eigenvalues of  \eqref{output} as a function of the parameters
$|a|$ and $\vp$, we use the characteristic polynomial of the
output density matrix to estimate the effect of small changes
after reduction to a 3-dimensional problem.

Since  $ \tfrac{1}{4} (1- \mu)(1- \lambda^2)$ is clearly an
eigenvalue of  \eqref{output}, we are left with the eigenvalue
problem for the $3 \times 3$ matrix
\begin{footnotesize}
  \be   \label{alt4}
&&\Delta =     (1 \pp \lambda^2) I +  \\
&&\pmx    c_\vp \,\lambda S +  M_ {\mu}|a|^2 &
     C+ i  s_\vp    \,   \lambda S& M_ {\mu}\sqrt{ |a|^2(1 - |a|^2) } \\
       C- i  s_\vp    \,   \lambda S&
         -c_\vp   \,\lambda S  & 0   \\
    M_ {\mu}  \sqrt{ |a|^2(1 - |a|^2) } &    0 &     -2 \lambda^2 +  M(1 - |a|^2)  \emx  \nn \ee
    \end{footnotesize}
where $C, S, \vp$ and $M_\mu$ are defined as in \eqref{output}.

The characteristic polynomial for $  \Delta - ( 1 + \lambda^2)I $
is \be  \label{poly}
 \wh{R}(\zeta)  & = &    -\zeta^3 +  (M_ {\mu}\mm  2 \lambda^2) \zeta^2 +
           \big[C^2 \pp \lambda^2 S^2  \pp  M_ {\mu}|a|^2 ( 2 \lambda^2 \nn \\
          & \pp &  c_{\vp}  \lambda S )] \zeta
          +  (C^2 \pp \lambda^2 S^2)(2 \lambda^2 - M)   \nn \\
          &+& M_ {\mu}|a|^2  \left( C^2 \pp \lambda^2 S^2 \pp 2
  \lambda^3 c_\vp  S \right)  \nn \\
    & \equiv &   R_0(\zeta) + M_ {\mu}|a|^2 ( 2 \lambda^2 +  c_{\vp}  \lambda S) \zeta \nn \\
    &+&  M_ {\mu}|a|^2  \big(C^2 + \lambda^2 S^2 + 2   \lambda^3 c_\vp  S
    \big) \; .
    \ee
This defines $R_0(\zeta) $. Then with $x = \zeta + 1 + \lambda^2$,
the  characteristic polynomial  for $ \Delta$  is
 \be
 \label{Rphi}
  R(x) & =& R_0(x - 1 - \lambda^2) + M_ {\mu}|a|^2
     \Big[ \big( 2 \lambda^2  \pp c_\vp \lambda S  \big)  x  \nn
   \\   &+&
        (1-\lambda^2)\big(2 \lambda^2 - c_\vp \lambda S  - S^2
        \big)
        \Big] \; ,
 \ee
where we used $C^2 + S^2 = 4 \lambda^2$.

\subsection{Optimization for  $ 1 < p < 2$}   \label{sect:1p2}

We first assume that $a, S$ are fixed and consider the optimal
value of $c_\vp = \cos \vp$ by examining the effect of the change
$c_\vp \mapsto c_\vp + \e$.  We apply Lemma~\ref{lemma:root}  with
$P(x) = R(x)$ given by \eqref{Rphi}, and $\delta_1 = \e M_
{\mu}|a|^2  \lambda S$ and $ \delta_2 = - \e M_ {\mu}|a|^2
\lambda S
  (1 - \lambda^2)$ to find that the eigenvalues of $\Delta$
  \be
 v_k  \mapsto   v_k + \e \, M_ {\mu}|a|^2  \lambda S  \frac{1}{(v_k
- v_m)(v_k - v_n)}  [v_k  - (1 - \lambda^2)] \nn \\ \label{evch1}
  \ee
with $k,m,n$ distinct. It then follows from Lemmas~\ref{lemma:p}
and \ref{lemma:q} using the expressions \eqref{deriv1} and
\eqref{deriv2} that
  \be   \label{normch2}
    \norm{ {\bf v}}^p \mapsto   \norm{ {\bf v}}^p +  \e M_ {\mu}|a|^2  \lambda S  \frac{p}{v_1 \mm v_3}
       \Big[  p \, (\acute{v}_1^{p-1} \mm  \acute{v}_3^{p-1}) \nn \\
       -  (p \mm 1) (1 \mm  \lambda^2)
          (\grave{v}_1^{p-2} \mm  \grave{v}_3^{p-2} ) \Big] + O(\epsilon^2)
  \ee
where $v_1 > \acute{v}_1 > v_2 > \acute{v}_3 > v_3$ and similarly
for $\grave{v}_k$. Then the quantity in square brackets  $[ ~~ ] >
0$ is positive for $1 < p < 2$.   Since $\lambda S = 2 \lambda^2
\sin \theta \geq 0$ (by our assumption $\theta \in [0,
\tfrac{\pi}{2}]$), $\norm{ \Delta }_p^p$ increases  as $c_\vp $
goes from $-1$ to $+1$ and is thus largest when $c_\vp= 1$.

To study the effect of changing $|a|$, first observe that when $
c_{\vp} = 1$,  $2 \lambda^2 - c_\vp \lambda S  - S^2 = 2 \lambda^2
(1 +  \sin \theta )(1 - 2   \sin \theta )$. If we insert this in
\eqref{Rphi}, we can apply Lemma~\ref{lemma:root}  with $P(x) =
R(x)$, $\delta_1 =  \e 2 \lambda^2 M_ {\mu}   (1 +  \sin \theta)$
and $\delta_2 = \e 2 \lambda^2 M_ {\mu}   (1 +  \sin \theta)  (1 -
\lambda^2)(1 - 2   \sin \theta) $ to conclude that \be
\label{evch2}
     v_k  \mapsto v_k + \e \, 2 \lambda^2 M_ {\mu} (1 +  \sin \theta)  \frac{1}{(v_k - v_m)(v_k - v_n)}
              \big[ v_k \nn \\
              + (1 - \lambda^2)(1 - 2   \sin \theta) \big].
\ee
We again apply  Lemmas~\ref{lemma:p} and \ref{lemma:q} using
the expressions \eqref{deriv1} and \eqref{deriv2} to conclude that
  \be    \label{normch22}
   \norm{ {\bf v}}^p   \mapsto   \norm{ {\bf v}}^p +
          \e  B      \Big[  p \, (\acute{v}_1^{p-1} \mm  \acute{v}_3^{p-1}) +  (p \mm 1) (1 \mm  \lambda^2)
         \nn \\ (1 \mm 2   \sin \theta)  (\grave{v}_1^{p-2} \mm  \grave{v}_3^{p-2} ) \Big] + O(\epsilon^2)
  \; \; \; \; \; \ee
with $\acute{v}_k, \grave{v}_k$ constrained as above and $B  =  2
\lambda^2 M_ {\mu}  (1 +  \sin \theta)  \frac{p}{v_1 \mm v_3}  >
0$. However, $  \grave{v}_1^{p-2} < \grave{v}_3^{p-2} $ because $p
- 2 < 0$. Therefore, we can only conclude that the quantity in
brackets is positive when $ \sin \theta  \geq \half$. Otherwise,
\eqref{normch2} has two competing positive and negative terms.

When  $\mu >   \frac{1 - \lambda^2}{3 - \lambda^2}$, the optimum
in \eqref{thetaopt} satisfies  $ \sin \theta > \half$. Even if the
$p$-norm does not increase monotonically  with $|a|$ for small
values of $\theta$, it seems likely that the optimum is still
achieved when $|a| = 1$.

\subsection{Optimization for  $p > 2$ } \label{sect:p>2}

When $p > 2$, the expression \eqref{normch2} contains competing
terms and we cannot reach a definite conclusion about the effect
of $c_\vp \mapsto c_\vp + \e$. However, both  \eqref{evch1} and
\eqref{evch2}  still imply that the largest eigenvalue increases
under the changes   $c_\vp \mapsto c_\vp + \e$ and $|a|^2 \mapsto
|a|^2 + \e$ with $\e > 0$. Moreover, if one fixes $\vp$ and
considers the change $|a|^2 \mapsto |a|^2 +  (\delta_1 x +
\delta_2)$ with \be \delta_1  &=&   \e 2 M_\mu  \lambda^2 (1 +
c_\vp \sin \theta) x  \\
\delta_2   &=&
   \e 2 M_\mu \lambda^2  (1-\lambda^2) (1 -  c_\vp \sin \theta - 2 \sin^2 \theta)
\ee then Lemmas~\ref{lemma:root} and \ref{lemma:p} imply that \be
\label{normpbig1}
 \norm{ \Delta}_p^p & \mapsto &   \norm{ \Delta}_p^p + \e B
  p  (1 + c_\vp \sin \theta) (\acute{v}_1^{p-1} \mm  \acute{v}_3^{p-1})  \nn \\
 &+&  \e B  (p \mm 1)
         (1-\lambda^2) (1 -  c_\vp \sin \theta \nn \\ &-& 2 \sin^2 \theta)
           (\grave{v}_1^{p-2} \mm  \grave{v}_3^{p-2} )  + O(\epsilon^2)
\ee
where $B = 2 \lambda^2 M_ {\mu}p$.    When $ \sin \theta <
\half$ this implies that $ \norm{ \Delta}_p$ increases with
$|a|^2$.   This is sufficient to show that when $  \mu <   \frac{1
- \lambda^2}{3 - \lambda^2} $ at least a  local optimum is
achieved when $c_\vp = 1$ and $\theta$ is given by
\eqref{thetaopt}.

\subsection{Additional heuristics}  \label{sect:heurp}

One can apply the  MVT  again to \eqref{normch22} using $ \acute{
v}_2 , \grave{v}_2$  to denote the mean values. Under the
assumptions  $ \acute{ v}_k  \approx \grave{v_k}  $ and $ \acute{
v}_2  \geq  (1 - \lambda^2)   $ the term  in square brackets
$[~~]$ in \eqref{normch22}  is
  \be
     \approx (p-1)  \tfrac{\acute{v}_1 -  \acute{v}_3}{\acute{ v}_2^2} (1 - \lambda^2) 2
           \big[ (p-1)(1-   \sin \theta)  +   \sin \theta \, \big] ~ \geq ~
           0 \nn \\
  \label{normch2b} \ee
when $1 < p < 2$. For $p > 2$, a similar analysis  beginning from
\eqref{normpbig1} gives a change in $ \norm{ \Delta}_p^p$
approximately proportional to
 \be
   (p-2) (1 - \sin^2 \theta) + (1 + c_\vp \sin \theta)  ~ \geq ~ 0 .
 \ee
This approach can even be applied to the  entropy to show that
when  $|a|$ increases the   output entropy decreases under the
assumptions above. We omit the details.

Since $v_1 >  \acute{v}_1 > \acute{v}_2 > \acute{v}_3 > v_3$,  one
knows that $\acute{v}_2$ is larger than the smallest eigenvalue of
$\Delta$ and probably close to the second largest, which one
expects to be $> 1- \lambda^2$. Thus, the assumptions above are
reasonable.  However, this is far from the desired proof that the
output $p$-norm increases with $|a|$ and is thus optimal for $|a|
= 1$. For  $p > 2$,  these heuristics are more convincing because
errors from these approximations are better controlled .

\section{Conclusions}    \label{sect:conc}

\subsection{Main results}

\begin{conj}   \label{conj:main}
Let $  \Phi_{\beta_0,\mu,\lambda}$ be a channel  on $M_4$
as defined in  \eqref{genchan} with $\mu, \lambda \in (0,1)$ and
 let  $\mu_c =  \frac{1 - \lambda^2}{2 - \lambda^2}$.
 Then

{\rm i)}  For  $0 < \mu < \mu_c$ the maximal output $p$-norm  is
achieved with a family of  input states of the form  $ V^T \ot  V
| \psi_\theta \ket $ with $V$ unitary and $ | \psi_\theta \ket $
given by \eqref{input} with $\theta = sin^{-1} \big( \frac{\mu }{1
- \mu} \frac{1}{1 - \lambda^2} \big) \in (0,\frac{\pi}{2})$.

{\rm ii)}   For  $\mu \geq \mu_c $  the optimal output $2$-norm
is achieved if and only if the input is the maximally entangled
state $|\beta_0 \ket $.

\noindent Moreover, the same conclusions hold for the minimal
output entropy.
\end{conj}

In Section~\ref{sect:p2} this conjecture was proved for $p = 2$.
Since $\norm{\gamma}_\infty$ is simply the largest eigenvalue of
$\gamma$,  the observations at the start of
Section~\ref{sect:p>2}, imply that the optimal input is achieved
for $a = 1$ when $p = \infty$.  The conjecture then follows from
the results in Section~\ref{sect:a1opt}.
   \begin{thm}
   Conjecture~\ref{conj:main} holds for $p = 2$ and for $p = \infty$.
   \end{thm}

Conjecture~\ref{conj:main} is supported by extensive numerical
work, as discussed in Section~\ref{sect:num}. Additional evidence
for the conjecture can be summarized as follows.

   \begin{itemize}

\item[a)]   When  $1 < p < 2$ and $\mu >  \frac{1 - \lambda^2}{3 -
\lambda^2}$ either  Theorem~\ref{conj:main} holds or the optimal
input is achieved with a state of the form $ U \ot  V  |
\psi_\theta \ket $ with $U \neq V^T$ and $0 < \theta < \sin^{-1}
\half$. The latter would be  unexpected, but has not been
excluded.

\item[ b)]   When  $p > 2$ and $\mu <  \frac{1 - \lambda^2}{3 -
\lambda^2}$ either  Theorem~\ref{conj:main} holds or the optimal
input is achieved with a state of the form $ U \ot  V  |
\psi_\theta \ket $ with $U \neq V^T$ and $  \theta > \sin^{-1}
\half$. Again, this seems unlikely.

\item[ c)]  The heuristic argument described in
Section~\ref{sect:heurp} makes other behavior  unlikely,
especially for $p > 2$.

\item[ d)]  For $1 < p < 2$, Corollary~\ref{cor:notmax}   below
excludes the possibility that the optimal input is maximally
entangled for $\mu < \mu_c$.

  \end{itemize}

The conjectured behavior is very different from that for the
shifted depolarizing channel on $M_4$ which maps
   \be
     R \mapsto  (1-\mu)\big[ \lambda R + (1-\lambda) (\tr R) \tfrac{1}{4} I_4 \big] + \mu (\tr R) \proj{\beta}
   \ee
where $|\beta \ket$ is fixed and the quantity in square brackets
$[ ~] $ is easily recognized as the usual depolarizing channel on
$M_4$.  In this case the optimal input is always achieved using
the vector $| \beta \ket $ which defines the shift.    Changing
from the usual depolarizing channel on $M_4$ to a product of qubit
channels $\Psi_\lambda \ot \Psi_\lambda$ dramatically changes the
effect of the correlation introduced by a maximally entangled
shift  $| \beta \ket $ for values of $\mu < \mu_c$.

Even without proving the conjecture, the next theorem implies that
for $1 < p  \leq  2$ and for the minimal output entropy, the
optimal input is never achieved with a maximally entangled state
when $\mu < \mu_c$. Moreover, to extend this result to $p  > 2$,
it would suffice to show that the optimal output is achieved with
$c_\vp = 1$.
       \begin{thm}    \label{max}
For $1 < p \leq 2$, the maximal output $p$-norm of the channel $
\Phi_{\beta_0,\mu,\lambda}$ defined in  \eqref{genchan}  is
achieved with a maximally entangled input if and only if that
input is $|\beta_0 \ket $. The same result holds for the minimal
output entropy.
   \end{thm}
\pf for $1 < p \leq  2$ the optimal output is  achieved with
$c_\vp = 1$. For any $p$ the optimal input is maximally entangled
if and only if $ \theta = \frac{\pi}{2}$ in \eqref{theta}. Then
$\sin \theta = 1$ and \eqref{alt4}  implies
  \be
      && \Delta - (1 + \lambda^2) I \nn \\
       &=&  \pmx    +  M_ {\mu}|a|^2 & 0 & M_ {\mu}\sqrt{ |a|^2(1 - |a|^2) } \\
     0  & -  2 \lambda^2  & 0   \\
    M_ {\mu}  \sqrt{ |a|^2(1 - |a|^2) } &    0 &     -2 \lambda^2 +  M(1 - |a|^2)  \emx
  \nn \\ \ee
from which it follows immediately that one eigenvalue of $\Delta$
is $1 + \lambda^2 - 2 \lambda^2 = 1 - \lambda^2$ and the other two
are
   \be
    1 + \lambda^2 +\half M_ {\mu}\pm \sqrt{ \frth M_{\mu}^2 + 4 \lambda^4 -  2 \lambda^2 M_ {\mu}(1 - 2 |a|^2) }
 \ee
which gives the largest $p$-norm when the term under the square
root is largest, i.e., when $|a|^2 = 1$.   But then, the results
of Section~\ref{sect:a1opt} apply.
   \qquad \qed
 \begin{cor}  \label{cor:notmax}
For $1 < p < 2$ and $\mu < \mu_c$ the optimal output of the
channel $  \Phi_{\beta_0,\mu,\lambda}$ is never attained with a
 maximally entangled input.   The same result holds for the minimal output entropy.
 \end{cor}

\subsection{Majorization  and trumping} \label{sect:trump}

Let  $|\psi_{\rm opt} \ket $ be the canonical vector $|\psi_\theta
\ket$ for  the family of input states which achieves the optimal
2-norm output for $\Phi_{\beta_0,\lambda,\mu}$, or, equivalently a
state of the form $V \ot V^T |\psi_\theta \ket$ as described in
part (i) of Conjecture~\ref{conj:main}.   If true, this conjecture
would imply that for any $ p > 1$, the output $p$-norms satisfy
\be \label{psiopt}
    \norm{ \Phi_{\beta_0,\lambda,\mu}\big( \proj{ \psi_{\rm opt}  } \big) }_p
      & \geq &    \norm{ \Phi_{\beta_0,\lambda,\mu}\big( \proj{ \psi } \big) }_p
\ee
for any other input state $|\psi \ket$. One might, therefore,
expect that the eigenvalues of the matrix
$\Phi_{\beta_0,\lambda,\mu}\big( \proj{ \psi_{\rm opt}  } \big)$
majorize those of any other output.   This is false, as can be
seen from the following two examples.

First, let  $\lambda= \tfrac{1}{3}$ and $\mu=\half$. In this case
$\mu>\mu_c = \tfrac{8}{17}$ so that the optimal input   is
$|\beta_0\rangle$, whose output eigenvalues are $\{
0,667,0.111,0.111,0.111 \}$.   A numerical search found that the
input product state of the form~(\ref{psiform}) with $a_0=a_2=0$
and $a_3 =  \1rt2 = -i a_1$ yields an output state with
eigenvalues $\{ 0.611,0.222, 0.111, 0.056 \}$ which is clearly not
majorized by those of the preceding state. For an example with
$\mu < \mu_c$, consider $\lambda= \half$ and $\mu= \tfrac{1}{4}$.
In this case the optimal input  gives an output with eigenvalues
$\{0.596,0.141, 0.141, 0.123 \}$. However, the same product  state
now yields an output with eigenvalues $\{0.422, 0.391, 0.141,
0.047 \}$ which, again, are not majorized by those of the
preceding state.

However, weak majorization  $x \prec_w y $ (which does not require
equal $1$ norms) is well-known to be equivalent to the stronger
condition that $\norm{x} \leq \norm{y}$ for all unitarily
invariant norms  \cite[Corollary 3.5.9]{HJ2}. This is known as the
Ky Fan dominance theorem \cite{Fan}.   Recently, Aubrun and
Nechita \cite{AN1,AN2} proved an analogous result for $\ell_p$
norms with $p \geq 1$ in which majorization is replaced by a
relation known as ``trumping''.    This concept, introduced in
connection with the phenomenon known as entanglement catalysis
\cite{JP} plays an important role in quantum information theory.
If there is a vector $z$ such that $x \ot z \prec y \ot z$, then
one says that the vector $y$  trumps  $x$  and writes $x \trump y
$.   Moreover, it follows that $\norm{x}_p \leq \norm{y}_p$ for
all $p > 1$. Aubrun and Nechita \cite{AN1,AN2} showed that
$\norm{x}_p \leq \norm{y}_p$  for all $p > 1$ if and only if $x$
is in the closure of the vectors trumped by $y$. Here, the closure
is taken in the $\ell_1$ norm and includes arbitrarily large
catalyst vectors $z$.   For a precise statement and other
formulations, see \cite{AN1,AN2}.

This leads us to the following reformulation of
Conjecture~\ref{conj:main} \begin{conj}   \label{conj:trump} Let $
\Phi_{\beta_0,\mu,\lambda}$ be a channel  on $M_4$ as defined in
\eqref{genchan} and let  $\mu_c =  \frac{1 - \lambda^2}{2 -
\lambda^2}$.

{\rm i)}  For  $0 < \mu < \mu_c$, let  $\psi_{\rm opt}  $ be as in
\eqref{psiopt}. Then if  $| \psi \ket  \neq (V^T \ot V) |\psi_{\rm
opt} \ket $ for some unitary $V$, the eigenvalues  of
$\Phi_{\beta_0,\lambda,\mu}\big( \proj{ \psi } \big)$ yield a
vector in the closure of the set of vectors trumped by the
eigenvalues of $   \Phi_{\beta_0,\lambda,\mu}\big( \proj{
\psi_{\rm opt}  } \big)$.

{\rm ii)}  For  $\mu \geq \mu_c $  and any input   $| \psi \ket
\neq |\beta_0 \ket $ the eigenvalues  of
$\Phi_{\beta_0,\lambda,\mu}\big( \proj{ \psi } \big)$ yield a
vector in the closure of the set of vectors trumped by the
eigenvalues of $   \Phi_{\beta_0,\lambda,\mu}\big( \proj{ \beta_0
} \big)$.

\end{conj}

Conjecture~\ref{conj:main} would follow immediately from
Conjecture~\ref{conj:trump}. It is, therefore, tempting to seek an
analytic proof of this conjecture by  seeking a catalyst $z$.
Unfortunately, this is not easy in general.   In our case, the
fact that $(1-\mu)(1-\lambda^2)$ is always an output eigenvalue,
allowed a reduction to an effective $3$-dimensional problem.   For
$d = 3$, it is known \cite{JP} that trumping with a finite
dimensional catalyst $z$ is never possible, although examples are
known  \cite{Daf,DK} which use a infinite-dimensional catalyst.
Thus, although we believe Conjecture~\ref{conj:trump} holds, it
seems more likely to be established by proving
Conjecture~\ref{conj:main}  and then applying Aubrun and Nechita's
results \cite{AN1,AN2} than by finding a catalyst.

We can make additional reformulations suggested by the arguments
in Section~\ref{sect:heur}. For example, let  $\Delta$ be given by
\eqref{alt4} with $\mu, \lambda, \theta, \vp $ fixed, and let
$\bv(|a|)$ denote its eigenvalues. Then  if $\bv(|a|) \trump
\bv(|1|)$ we could conclude that the optimal $p$-norm of
$\Phi_{\beta_0, \mu,\lambda}$  is attained when $|a| = 1$. The
validity of  Conjecture~\ref{conj:main} would then follow from the
arguments in Section~\ref{sect:a1opt}.

\bigskip

\noindent{\bf Acknowledgment}  It is a pleasure to recall that
this work began when MBR and VG were visiting the  Quantum
Information Theory Group at the University of Pavia. The authors
VG, CM and MBR also benefitted from discussions during workshops
in Benasque and at the ICTP in Trieste. The contribution of MBR
was partially supported by the National Science Foundation under
Grants DMS-0314228 and DMS-0604900, and by the National Security
Agency and Advanced Research and Development Activity under Army
Research Office contract number DAAD19-02-1-0065. The contribution
of FC and VG has been in part supported by the Centro di Ricerca
Ennio De Giorgi of the Scuola Normale Superiore.

\bigskip

 \appendix

\section{Entanglement parametrization}
        \label{app:entang}

Let $| \beta_0 \ket = \1rt2 (|00 \ket + |11 \ket)$, and define $|
\beta_k  \ket = (I \ot \sigma_k)| \beta_0 \ket$ for $k = 1,2,3$
and $\sigma_k$ denotes the Pauli matrices. These  four orthogonal
maximally entangled states form an orthonormal basis for ${\bf
C}^4$. The property $   \trp_1 |\beta_k \kb \beta_m | = \half
\sigma_k \sigma_m   $,  facilitates computation of reduced density
matrices and entanglement of pure states represented in this
basis.  Our notation differs  by a factor of $i$ from the
so-called ``magic basis'' introduced in \cite{BDSW}.

When $\sigma_k$ acts on the first qubit,
\be  \label{bellleft}
   ( \sigma_k \ot I )| \beta_0 \ket = \begin{cases} | \beta_k \ket & k =1,3 \\
                  -  | \beta_2 \ket & k = 2 \end{cases}
                  \ee
An arbitrary pure state  $|\psi \ket \in {\bf C}_4$ can be written
in the form \be  \label{psiform}
   |\psi \ket & = &    \sum_{k = 0}^3  a_k  | \beta_k  \ket  \, = \,
         (I \ot  A )  | \beta_0  \ket
\ee where $A =   \sum_k a_k  \sigma_k$ is in $M_2$. This gives an
isomorphism between vectors in $ {\bf C}_4$  and matrices in
$M_2$. We henceforth restrict ourselves to normalized vectors for
which $\norm{ \psi }^2 =\sum_{k = 0}^3 |a_k|^2 = 1 =  \half \tr
A^\dag A$. Combining \eqref{bellleft} with the observation that
$a_2 \mapsto - a_2$ takes $A \mapsto A^T$, we recover the
well-known result   that
 \be  \label{trans}
      |\psi \ket  =  (I \ot A) |\beta_0 \ket = (A^T \ot I)  |\beta_0 \ket
\ee

Since local unitary transformations do not affect the entanglement
of $|\psi \ket$, the state in \eqref{psiform} has the same
entanglement as any state of the form \be   \label{SVD}
     (U \ot V) |\psi \ket & = &  \sum_k a_k (U \ot I) (I \ot V) (I \ot \sigma_k )   | \beta_0  \ket  \nn \\
                & = &   (I \ot V)   \sum_k a_k (I \ot \sigma_k )  (I \ot U^T)  | \beta_0  \ket  \nn \\
                & = &   (I \ot  VA U^T)   | \beta_0  \ket .
\ee
Moreover, by the singular value decomposition (SVD), one can
choose $U,V$ so that $VA U^T$ is diagonal and positive, i.e., $VA
U^T  = D = a_0 I + a_3 \sigma_3 \geq 0$. By including a suitable
permutation in $U,V$ one can further require that the singular
values are in decreasing order, which is equivalent to $a_0 \geq
a_3 \geq 0$. Thus, we can choose $U,V$ unitary so that \be
\label{psiUV}
    |\wh{\psi}\ket  ~ = ~  (U \ot V) |\psi \ket ~ = ~    (I \ot  VA U^T)  & = &
                 a_0 |\beta_0 \ket + a_3  |\beta_3 \ket    \label{psiUVang}
\ee with $a_0 \geq  a_3 \geq 0$ and $a_0^2 = a_3^2 = 1$.    When
this is rewritten as in \eqref{theta},  this implies that $\cos
\tfrac{\theta}{2} \geq  \sin \tfrac{\theta}{2} \geq 0$ which
implies $ \tfrac{\theta}{2} \in [0, \tfrac{\pi}{4}]$ and hence
$\theta \in  [0, \tfrac{\pi}{2}]$. We emphasize that \eqref{psiUV}
is {\em not} the most general pure state of this ``diagonal''
form; instead $ |\wh{\psi}\ket $ should be viewed as a canonical
state from which all others with the same entanglement can be
written as $(U^\dag \ot V^\dag)  |\wh{\psi}\ket $. We use the same
notation as in \eqref{key} because the equivalence class of states
defined by \eqref{enteq} corresponds exactly to the equivalence
class of matrices with the same singular values.

In the form \eqref{psiUV}   it is easy to find the entanglement of
$ |\psi \ket $ because \be
    \trp_1 |\beta_0 \kb \beta_3 | =  \trp_1 \proj{\beta_0}  (I \ot \sigma_3) =
         ( \half I_2) \sigma_3 =   \half \sigma_3.
\ee
Therefore, the reduced density matrix of \eqref{psiUV} is
\be   \label{reddm}
       \gamma =   \half (I + 2 a_0 a_3  \, \sigma_3 ) \;,
             \ee
and its entanglement  is  $ S(\gamma) = h(2 a_0 a_3)$
where $h(x)$ is the binary entropy
\be  \label{hdef}
    h(x) \equiv  - \tfrac{1+x}{2} \ln \tfrac{1+x}{2} - \tfrac{1-x}{2} \ln \tfrac{1-x}{2} \;.
 \ee

It this paper we consider  subfamilies of states of the form
\eqref{psiUV} with  $UV^T$ fixed.    When  $UV^T = I$, $ U =
\ovb{V}$ and the SVD becomes $VA V^\dag$, which is the standard
form for diagonalizing a self-adjoint matrix. Note that $A$ is
self-adjoint if and only if all $a_k$ in \eqref{psiform} are real
and this property is preserved by the transformation $VA V^\dag$.
Thus, if we begin with a state  $ | \wh{\psi} \ket  $  of the
diagonal form \eqref{psiUV} with matrix $D = a_0 I + a_3
\sigma_3$, then the family of states \be
     |\psi \ket  = (V^T \ot V^\dag) | \wh{ \psi} \ket =
           (I \ot V^\dag D V ) |\beta_0 \ket = (I \ot A) |\beta_0 \ket
\ee
 with $V$ unitary all have real coefficients $a_k$ when written in the form \eqref{psiform}.

Since any maximally entangled state can be written as $|\beta \ket
=  (U_1 \ot U_2)  | \beta_0 \ket  = {(I \ot U_2 U_1^T)  | \beta_0
\ket } $, the vector $ | \beta_0 \ket  $ can be regarded as the
canonical representative of the equivalence class of maximally
entangled states. This also gives a one-to-one correspondence
between maximally entangled states and unitary matrices in $M_2$
via the relation
 \be
 \label{maxentU}
     |\beta \ket & =  &(I \ot U)  | \beta_0 \ket    =   (U^T \ot I)        | \beta_0 \ket .
  \ee
A matrix is unitary if and only if it can be written (up to an
overall phase factor) as $U = u_0 I + i \sum_k u_k \sigma_k$ with
$u_k$ real. Thus, a state written in the form \eqref{psiform} with
$a_0$ real is maximally entangled if and only if  ${\rm Re} \, a_k
= 0 $ for $k = 1,2,3$.

\section{Comparison lemmas } \label{app3}

For $r_1 > r_2 > r_3$ it is elementary to  verify that
\begin{footnotesize}
\be
    \frac{1}{ (r_1 - r_2)(r_1 - r_3)} +   \frac{1}{ (r_2 - r_1)(r_2 - r_3)}  +
       \frac{1}{ (r_3 - r_1)(r_3 - r_2)}  ~ = ~  0 \nn \\ \label{sum0a}
\ee and \be
    \frac{r_1}{ (r_1 - r_2)(r_1 - r_3)} +   \frac{r_2}{ (r_2 - r_1)(r_2 - r_3)}  +
       \frac{r_3}{ (r_3 - r_1)(r_3 - r_2)}   ~ = ~  0 \nn \\
\label{sum0b} \ee \end{footnotesize}
\begin{lemma}   \label{lemma:root}
Let  $P(x) = - (x-r_1)(x - r_2)(x - r_3)$ be the cubic polynomial
with real roots $r_1 >  r_2 >  r_3 $ and let $Q(x) = P(x) +
\delta_1 x + \delta_2$ with $\delta_1, \delta_2 > 0$.    Then the
roots of $Q(x)$ are approximately
 $s_k \approx  r_k + \tfrac{1}{(r_m - r_k)(r_n - r_k)}( r_k \delta_1 + \delta_2) $
 with $k,m,n$ distinct.
\end{lemma}
\pf  Near each of the roots, the tangent to $P(x)$ is
\be
 y = (x - r_k) P^{\prime}(r_k)  =  - (x  - r_k)  \tfrac{1}{ (r_m - r_k)(r_n - r_k)}
\ee
from which it follows that near $x = r_k$
\be
     Q(x) \approx  - (x - r_k) (r_m - r_k)(r_n - r_k)  + \delta_1 x + \delta_2.
\ee
It then follows that  $Q(s_k ) \approx 0$ if
\be
    s_k & = &  r_k  \frac{ 1}{ 1  - \frac{\delta_1}{ (r_m - r_k)(r_n - r_k) - \delta_1 } } +
       \frac{\delta_2}{ (r_m - r_k)(r_n - r_k) - \delta_1}  \nn  \\
       & \approx &  r_k + r_k  \frac{ \delta_1}{ (r_m - r_k)(r_n - r_k) } \nn \\
       &+&   \frac{\delta_2}{ (r_m - r_k)(r_n - r_k) - \delta_1} . \; \; \qed
\ee

\begin{lemma}   \label{lemma:p}
Let  ${\bf v}$ be a vector in ${\bf R}_3$ with all $v_1 > v_2 >
v_3  > 0$ and let $v_k \mapsto  w_k \equiv  \linebreak v_k +
\tfrac{1}{(v_m - v_k)(v_n - v_k)} \epsilon v_k $.   Then $\norm{
\bf w}_1 = \norm{ \bf v}_1 $ and $\norm{ \bf w}_p > \norm{ \bf
v}_p $ when $ \epsilon > 0$ and $p > 1$.
\end{lemma}
\pf  The fact that $\norm{ \bf w}_1 = \norm{ \bf v}_1 $ follows
from \eqref{sum0b}.   For $p > 1$, observe that
   \be
   \sum_k w_k^p  =  \sum_k v_k^p + p \epsilon v_k^{p-1}   \tfrac{1}{(v_m - v_k)(v_n - v_k)} \epsilon v_k
   +   O(\epsilon^2)
   \ee
Thus, up to $O(\epsilon^2) $,
 \begin{footnotesize}
 \be \norm{ \bf w}_p^p - \norm{ \bf
v}_p^p  &=& \epsilon  p \Big( \frac{v_1^p}{(v_1 - v_2)(v_1 - v_3)}
   -   \frac{v_2^p}{(v_1 - v_2)(v_2 - v_3)} \nn
   \\ &+&
    \frac{v_3^p}{(v_1 - v_3)(v_2 - v_3)} \Big)  \nn \\
    & ~ & \nn \\
    & = &    \epsilon  \, p \, \frac{1}{v_1 - v_3}  \Big( \frac{v_1^p - v_2^p }{ v_1 - v_2 } -
            \frac{v_2^p - v_3^p }{ v_2 - v_3 } \Big) ~ \geq ~ 0      \label{diff1} \\ \label{deriv1}
            & = &  \epsilon  \, p^2 \, \frac{1}{v_1 - v_3} \big(\acute{v}_1^{p-1} - \acute{v}_3^{p-1} \big) ~ \geq ~ 0
\ee
 \end{footnotesize}
where we used \eqref{sum0a}, and then the mean value theorem to
obtain \eqref{deriv1} with $v_1 \geq  \acute{v}_1 \geq v_2$ and
$v_2 \geq  \acute{v}_3 \geq v_3$. The inequality on the right in
\eqref{diff1} follows from the fact that $f(x) = x^p$ is convex
for $p > 1$. Although this suffices  to prove the Lemma, the
expression \eqref{deriv1}, will be useful when we need to compare
competing terms.   \qquad \qed

\begin{lemma}   \label{lemma:q}
Let  ${\bf v}$ be a vector in ${\bf R}_3$ with all $v_1 > v_2 >
v_3  > 0$ and let $v_k \mapsto  w_k \equiv  \linebreak v_k +
\tfrac{1}{(v_m - v_k)(v_n - v_k)} \epsilon $.   Then $\norm{ \bf
w}_1 = \norm{ \bf v}_1 $ and for $\epsilon > 0$ \bsq
\begin{align}
\norm{ \bf w}_p  & > \norm{ \bf v}_p   &\hbox{\rm for}  \quad &  p > 2   \qquad \\
\norm{ \bf w}_p  & < \norm{ \bf v}_p   &\hbox{\rm for}  \quad & 1 < p < 2   \qquad
\end{align}   \esq
\end{lemma}
\pf  The proof is identical to that of Lemma~\ref{lemma:p} above except
that \eqref{diff1} becomes
\be
\norm{ \bf w}_p^p - \norm{ \bf v}_p^p
    & = &    \epsilon  \, p \, \frac{1}{v_1 - v_3}  \Big( \frac{v_1^{p-1} - v_2^{p-1} }{ v_1 - v_2 } -
            \frac{v_2^{p-1} - v_3^{p-1} }{ v_2 - v_3 } \Big)   \nn \\ \label{deriv2}
             & = &  \epsilon  \, p(p-1) \, \frac{1}{v_1 - v_3} \big(\acute{v}_1^{p-2} - \acute{v}_3^{p-2} \big) \nn \\ \label{diff2}  \ee
where, as before $v_1 \geq  \acute{v}_1 \geq v_2$ and $v_2 \geq
\acute{v}_3 \geq v_3$. When $p >2$, the function $x^{p-1}$ is
convex; however, when $1 < p < 2$, it is concave.   Moreover, when
$p - 2 < 0$, the expression on the right in  \eqref{deriv2} is
negative because $ \acute{v}_3   \leq  v_2 \leq \acute{v}_1 $.
\qquad \qed

\begin{table}[h!]
\begin{footnotesize}
\bee  \begin{array}{ | l | l |} \hline  \quad \mu < \mu_c  & \quad
\mu \geq \mu_c \\ \hline
 \tfrac{1}{4}  (1- \mu)(1+ \lambda^2) + \tfrac{\mu}{2} + \half \sqrt{  \frac{\mu^2}{1 - \lambda^2}  + (1 - \mu)^2 \lambda^2} &
 \tfrac{1}{4}    (1- \mu)(1+ 3 \lambda^2) +   \mu \\
  \tfrac{1}{4}   (1- \mu)(1- \lambda^2) &    \tfrac{1}{4} (1- \mu)(1- \lambda^2) \\
     \tfrac{1}{4}  (1- \mu)(1- \lambda^2) &    \tfrac{1}{4} (1- \mu)(1- \lambda^2) \\
   \tfrac{1}{4}  (1- \mu)(1+ \lambda^2) + \tfrac{\mu}{2} - \half \sqrt{  \frac{\mu^2}{1 - \lambda^2}  + (1 - \mu)^2 \lambda^2}
   &   \tfrac{1}{4} (1- \mu)(1- \lambda^2)   \\
\hline \end{array} \eee
 \caption{Optimal output eigenvalues for $p = 2$} \label{tab:evopt}
\end{footnotesize}
\end{table}

\section{Eigenvalue comparison}

The eigenvalues of $\Delta$ are easy to find for the two boundary
cases of $a = 0 $ and $a = 1$. The lists in Table~\ref{tab:ev01}
are intended to be in decreasing order, but the two smallest
eigenvalues may  switch for very small $M_{\mu}$. Although the
largest eigenvalue is always greater for $a = 1$ than for $a = 0$,
the same is also true for the smallest.   This precludes using
majorization to conclude that all $p$-norms for  $a = 1$ exceeds
that for $a = 0$. However, as discussed in
Section~\ref{sect:trump}, the eigenvalues for $a=1$ could still
trump those for any $a < 1$. If so, the optimal output eigenvalues
for $p= 2$ given in Table~\ref{tab:evopt} are conjectured to also
be optimal for all $p > 1$.

\begin{widetext}
 \begin{table}[t!]     \label{tab:eval1}
\bee  \begin{array}{| lcl |} \hline \qquad a   = 0 & \quad  &
\qquad    a   = 1  \\   \hline 1 + \lambda^2 + \sqrt{ 4 \lambda^2
- (1- \lambda^2) S^2 }    & ~~   \nearrow  ~~ &
1 + \lambda^2 + \half M_ {\mu}+ \sqrt{ \frth M_{\mu}^2 + 4 \lambda^2 - (1- \lambda^2) S^2   + M_ {\mu} | \lambda S |  }  \\
   1- \lambda^2 + M_ {\mu}   &  \searrow &   1- \lambda^2\\
   1 + \lambda^2 - \sqrt{ 4 \lambda^2 - (1- \lambda^2) S^2 }
      &  \nearrow &  1 + \lambda^2 + \half M_ {\mu}-
   \sqrt{ \frth M_{\mu}^2 + 4 \lambda^2 - (1- \lambda^2) S^2   +  M_ {\mu} | \lambda S | }  \\ \hline
\end{array} \eee
\caption{Eigenvalues of $\Delta$ with arrows showing expected
increase and decrease with $a$.} \label{tab:ev01}
\end{table}
\end{widetext}

\end{document}